\documentclass[a4paper,11pt]{article}
\usepackage{pos}

\newcommand{\Sig}{\pmb{\sigma}}
\newcommand{\Del}{\pmb{\Delta}}
\newcommand{\Partial}{\pmb{\partial}}

\title{Measurement-based quantum simulation of Abelian lattice gauge theories}

\author[a]{Hiroki Sukeno}
\author*[b]{Takuya Okuda}

\affiliation[a]{Department of Physics and Astronomy, State University of New York at Stony Brook\\
Stony Brook, NY 11794-3840, USA}

\affiliation[b]{Graduate School of Arts and Sciences, University of Tokyo\\
Komaba, Meguro-ku, Tokyo 153-8902, Japan}

\emailAdd{sukeno.hiroki@stonybrook.edu}
\emailAdd{takuya@hep1.c.u-tokyo.ac.jp}

\abstract{The digital quantum simulation of lattice gauge theories is expected to become a major application of quantum computers. Measurement-based quantum computation is a widely studied competitor of the standard circuit-based approach. We formulate a measurement-based scheme to perform the quantum simulation of Abelian lattice gauge theories in general dimensions. The scheme uses an entangled resource state that is tailored for the purpose of gauge theory simulation and reflects the spacetime structure of the simulated theory. Sequential single-qubit measurements with the bases adapted according to the former measurement outcomes induce a deterministic Hamiltonian quantum simulation of the gauge theory on the boundary. 
We treat as our main example the $\mathbb{Z}_2$ lattice gauge theory in $2+1$ dimensions, simulated on a 3-dimensional cluster state.
Then we generalize the simulation scheme to Wegner's lattice models that involve higher-form Abelian gauge fields. 
The resource state has a symmetry-protected topological order with respect to generalized global symmetries that are related to the symmetries of the simulated gauge theories. 
We also propose a method to simulate the imaginary-time evolution with two-qubit measurements and post-selections.}

\FullConference{The 40th International Symposium on Lattice Field Theory (Lattice 2023)\\
July 31st - August 4th, 2023\\
Fermi National Accelerator Laboratory\\}

\begin{document}
\maketitle

\section{Introduction}

Euclidean lattice gauge theories~\cite{PhysRevD.10.2445} have been simulated on classical computers with a great success, even in
the non-perturbative parameter regime that is difficult to study analytically.
Still, there are simulation targets, such as real-time evolution and finite density QCD, for which the path integral formulation of lattice gauge theory suffers from the sign problem.
This is a difficulty in the evaluation of amplitudes due to the oscillatory contributions in the Monte-Carlo importance sampling \cite{PhysRevB.41.9301, 2008arXiv0808.2987H, 2017PTEP.2017c1D01G, 2021arXiv210812423N}.
In the Hamiltonian formulation, which is by construction free of the sign problem, the dimension of the Hilbert space grows exponentially with the size of the system. 
The quantum computer is expected to help us alleviating this curse of dimensionality, enabling us to simulate the quantum many-body dynamics in principle with resources linear in the system size \cite{feynman2018simulating, lloyd1996universal}. 
The quantum simulation of gauge theory is thus one of the primary targets for the application of 
quantum computers/simulators, whose studies are fueled by the recent advances in NISQ quantum technologies~\cite{2012JordanLeePreskill, Preskill:2018, 2016RPPh...79a4401Z,2013AnP...525..777W, 2016ConPh..57..388D,2020EPJD...74..165B, 2006PhRvA..73b2328B}.

\begin{figure}
    \centering
    \includegraphics[width=0.5\linewidth]{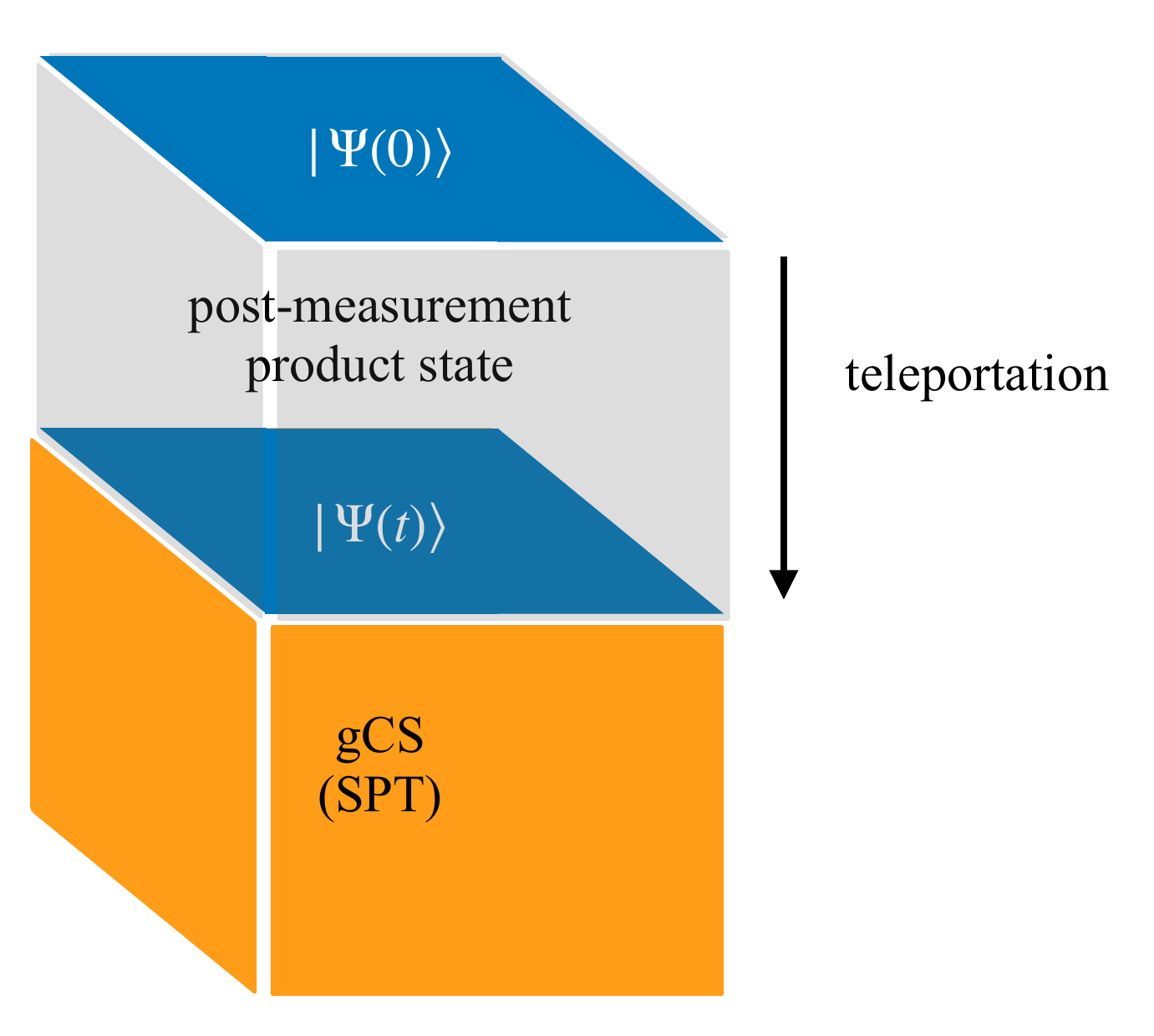}
    \caption{
 The schematic picture of MBQS.
 We begin by preparing a resource state with the initial wave function on the boundary. 
 By applying single-qubit measurements based on a measurement pattern, we obtain $|\Psi(t)\rangle$, the wave function after the evolution with the Hamiltonian of the gauge theory, at the boundary of the reduced lattice. }    \label{fig:concept}
\end{figure}

The goal of this contribution, based on~\cite{Sukeno:2022pmx}, is to present a new quantum simulation scheme for lattice gauge theories.
Our scheme, which we call {\it measurement-based quantum simulation} (MBQS), is motivated by the idea of measurement-based quantum computation (MBQC)~\cite{OneWayQC,Raussendorf2002ComputationalMU, PhysRevA.68.022312, briegel2009measurement, raussendorf2012quantum}.
In the MBQS of a gauge theory in $d$ spacetime dimensions, one prepares a resource state, which is obtained by entangling qubits located on various cells on a $d$-dimensional lattice.
We emphasize that our scheme uses a resource state that is tailored for the purpose of gauge theory simulation, so that the spatial structure of the resource state reflects the spacetime structure of the gauge theory.
The state of the gauge theory is in the Hilbert space of the qubits on the boundary.
By measuring qubits sequentially with the bases adapted according to the former measurement outcomes, one can perform the quantum simulation of the gauge theory deterministically.
The graphical representation of the MBQS scheme is shown in Figure~\ref{fig:concept}.

\section{Lattice models and resource states}\label{sec:Lattice models}

As a prototypical example, let us consider the gauge theory with gauge group $\mathbb{Z}_2 = \{0,1\text{ mod } 2\}$ on a two-dimensional square lattice.
We use the cell complex notation, where  $\Delta_0$ is the set of 0-cells (vertices) $\sigma_0$ , $\Delta_1$ the set of 1-cells (edges) $\sigma_1$, and $\Delta_2$ the set of 2-cells (faces) $\sigma_2$, and so on.
For a $k$-chain $c_k$, we write $Z(c_k) = \prod_{\sigma_k\in\Delta_k} Z_{\sigma_k}^{a(c_k;\sigma_k)}$, where $a(c_k;\sigma_k)$ are the coefficients in the expansion.
The Hamiltonian is given as
\begin{align}
H
&=-\sum_{\sigma_1 \in \Delta_1} X(\sigma_1) 
- \lambda \sum_{\sigma_2 \in \Delta_2}  Z(\partial \sigma_2) 
\ ,
\end{align}
where $\lambda$ is a coupling constant.
The 1-chain $\partial\sigma_2$ is the sum of the 1-cells $\sigma_1$ that are contained in the boundary of $\sigma_2$.
Physical states must satisfy the Gauss law constraint
\begin{equation}\label{eq:Gauss-law-Z2}
X(\partial^* \sigma_0) = 1
\end{equation}
in the absence of external charges, where $\partial^*$ is the boundary operator for the dual lattice and $\sigma_0$ is identified with a face in the dual lattice.
We wish to implement the Trotterized time evolution
\begin{align} \label{eq:trotter-step-of-3-2}
     T(t) &= 
    \Bigl( 
    \prod_{\sigma_1 \in \Delta_1}  e^{i  X(\sigma_1) \delta t }   \,
    \prod_{\sigma_2 \in \Delta_2} e^{i \lambda  Z(\partial \sigma_2) \delta t }
    \Bigr)^n 
\end{align}
with $t= n \delta t$. 

To realize $T(t) $ by measurements, we introduce the following resource state tailored to simulate the gauge theory.
Let us consider a 3-dimensional cubic lattice.
To differentiate 2- and 3-dimensional lattices, we use the bold font to denote objects associated with the latter.
We place qubits on 1-cells $\pmb{\sigma}_1 \in \pmb{\Delta}_1$ and 2-cells $\pmb{\sigma}_2 \in \pmb{\Delta}_2$.
The resource state (generalized cluster state) is obtained by applying controlled-Z gates to entangle the $+1$-eigenstates ($|+\rangle$) of the Pauli $X$ operator:
\begin{align}
|\text{gCS}\rangle = \left( \prod_{\pmb\sigma_1\subset \pmb\partial\pmb\sigma_2} CZ_{\pmb\sigma_1, \pmb\sigma_2}\right) |+\rangle^{\otimes \pmb\Delta_1\cup\pmb\Delta_2} \ .
\end{align}
This state is stabilized by the stabilizers
\begin{align}
K(\Sig_{n}) &= X(\Sig_n) Z(\Partial \Sig_n) \,,
\\
K(\Sig_{n-1}) &= X(\Sig_{n-1}) Z(\Partial^*\Sig_{n-1}) \,.
\end{align}

\begin{figure*}
    \centering
    \includegraphics[width=0.6\linewidth]{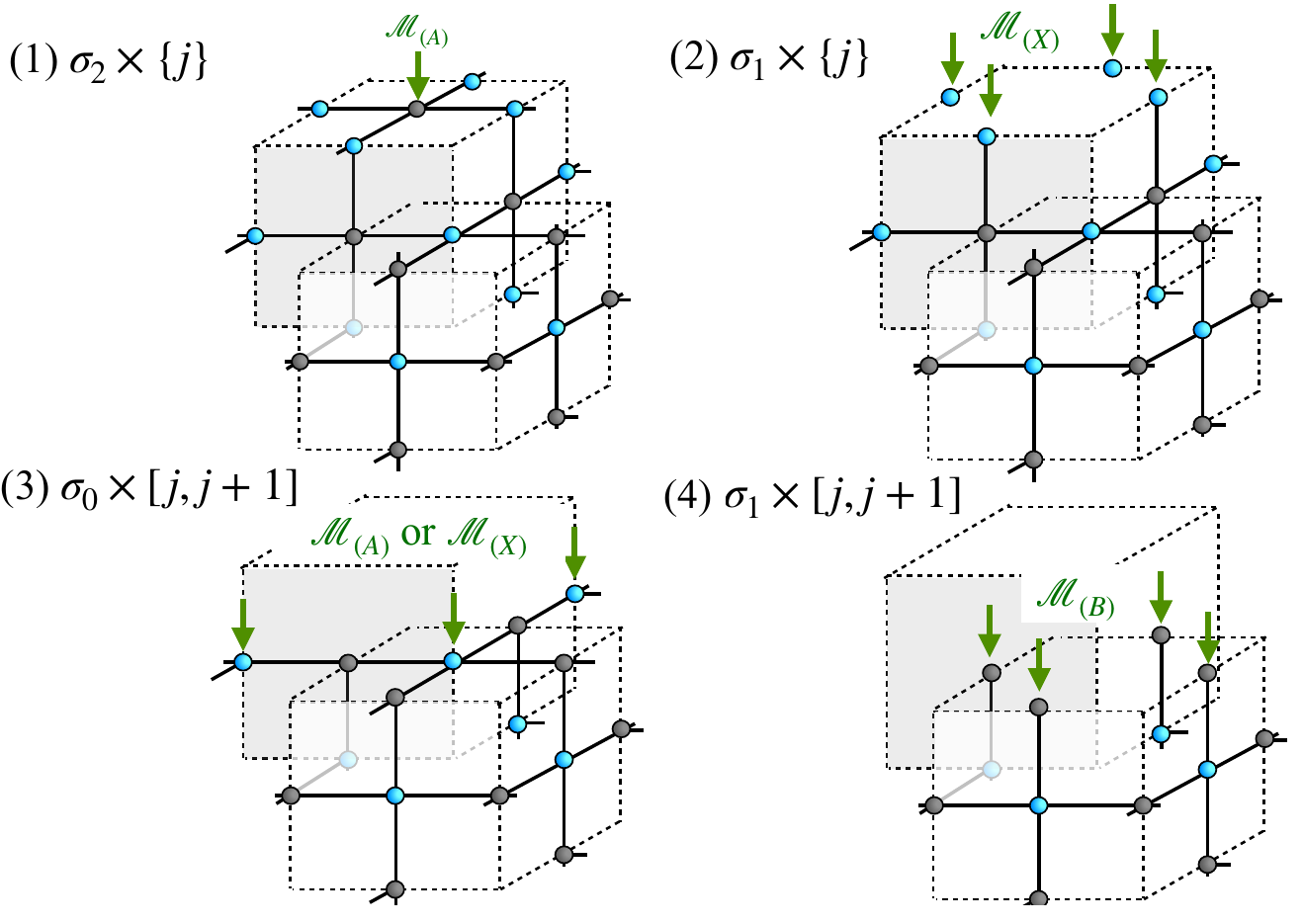}
    \caption{
    The protocol for the $\mathbb{Z}_2$ gauge theory in $2+1$ dimensions.
    The black balls represent the qubits living on 2-cells $\Sig_2 \in \Del_2$ and the blue ones are those on 1-cells $\Sig_1 \in \Del_1$. We use notations such that $\sigma_2\times\{j\}$ is a 2-cell in the 3-dimensional lattice at the vertical coordinate (which increases as as we go down vertially) $x_3=j$. Similarly, $[j,j+1]$ is an interval between $x_3=j$ and $x_3=j+1$, so that $\sigma_0\times[j,j+1]$ is a 1-cell in the 3-dimensional lattice.}
    \label{figure:RBH-pattern}
\end{figure*}

We indicate the measurement pattern for the simulation in Figure~\ref{figure:RBH-pattern} and in Table~\ref{table:z2-pattern}.
The measurement bases are defined as
\begin{align} \label{Wegner-ZZ-basis}
\mathcal{M}_{(A)}=
\left\{ e^{i\xi X}|s\rangle  \, \middle| \, s=0,1 \right\} \ .
\end{align}
and
\begin{align}
    \mathcal{M}_{(B)}= \left\{ 
    e^{i\xi Z} |\tilde{s} \rangle  \,\middle|\, s=0,1 \
    \right\} \ ,
\end{align}
where $|\tilde{s}\rangle$ is the eigenvector of the $X$ operator with the eigenvalue $(-1)^s$:
\begin{align}
    X|\tilde{s}\rangle = (-1)^s |\tilde{s}\rangle \ . 
\end{align}
The basis
$   \mathcal{M}_{(X)}=\{| \tilde{s} \rangle  \,|\, s=0,1\}$ is a specialization of $  \mathcal{M}_{(B)}$.
In~\cite{Sukeno:2022pmx} we show that the time evolution~\eqref{eq:trotter-step-of-3-2} is induced if the measurement angles $\xi$ are chosen adaptively based on the previous outcomes.
This is the main result of~\cite{Sukeno:2022pmx}.

\begin{table}
\begin{align}
\begin{array}{|r | c c c c c c c|}
    \hline
    \mbox{basis}
    & ~~~~ \mathcal{M}_{(A)} 
    &\rightarrow & \mathcal{M}_{(X)}
    & \rightarrow & \mathcal{M}_{(A)} 
    &\rightarrow & \mathcal{M}_{(B)} \\
    \mbox{layer}
    & ~~~~ pt
    & &   pt
    & &   I
    & &   I \\
    \mbox{3d cell} 
    & ~~~~ \pmb{\sigma_2}
    &  & \pmb{\sigma_1}  
    &  & \pmb{\sigma_1}
    &  & \pmb{\sigma_2}  \\
    \mbox{2d cell} 
    & ~~~~ \sigma_2  
    &  & \sigma_1  
    &  & \sigma_0 
    &  & \sigma_1  
    \\
    \hline
\end{array}
\nonumber
\end{align}
\caption{Measurement pattern for the $\mathbb{Z}_2$ gauge theory in $2+1$ dimensions.
$pt$ and $I$ denote a point and an interval in the vertical (fictitious time) direction.
The basis $\mathcal{M}_{(A)}$ in the third step induces an energy penalty for the violation of the Gauss law constraint (gauge invariance).
}
\label{table:z2-pattern}
\end{table}

The simulation protocol and the measurement pattern can be  generalized to models other than the $\mathbb{Z}_2$ gauge theory in $2+1$ dimensions.
A nice class of models that admits generalization is Wegner's models $M_{(d,n)}$~\cite{Wegner} that involve higher-form $\mathbb{Z}_2$ gauge fields in $d$ spacetime dimensions.
Another generalization we have achieved is to replace the gauge group $\mathbb{Z}_2$ to $\mathbb{Z}_N$ with $N>2$ or $\mathbb{R}$.
Further, we have generalized the protocol (measurement pattern) to the Majorana chain \cite{2001PhyU...44..131K}  and the imaginary-time evolution, and has established a correspondence between the statistical partition function and the resource state (generalizing  \cite{2008PhRvL.100k0501V, PhysRevLett.98.117207}).

\section{SPT order of the resource state}

The presence of an SPT order has been suggested to be an important ingredient of resource states for the ability to perform the (universal) MBQC \cite{ElseEtAlPhysRevLett.108.240505, ParakashWeiPhysRevA.92.022310, MillerMiyakePhysRevLett.114.120506, StephenPhysRevLett.119.010504,NautrupPhysRevA.92.052309, Miller2016, Chen:2018, WeiHuangPhysRevA.96.032317,PhysRevLett.122.090501, daniel2020computational, Devakul2018, 2022arXiv221005089R}.
In~\cite{Sukeno:2022pmx} we show that the resource state (generalized cluster state) gCS$_{(d,n)}$, constructed to simulate $M_{(d,n)}$ by measurements, has a symmetry protected topological order protected by global $(d-n)$- and $(n-1)$-form $\mathbb{Z}_2$ symmetries.
Whether the ability to simulate the relevant gauge theories is intrinsically related to the SPT order is left as an open question.

\section*{Acknowledgement}
HS was partially supported by the Materials Science and Engineering Divisions, Office of Basic Energy Sciences of the U.S. Department of Energy under Contract No. DESC0012704.
The work of TO is supported in part by the JSPS Grant-in-Aid for Scientific Research No. 21H05190.

\bibliographystyle{JHEP}
\bibliography{references}

\end{document}